\documentclass[%
reprint,
amsmath, amssymb,
aps,
prb,
10pt
]{revtex4-2}

\usepackage[cp1252, utf8]{inputenc}
\usepackage{verbatim}
\usepackage{standalone}
\usepackage{xcolor}
\usepackage{graphicx}
\usepackage{dcolumn}
\usepackage{dsfont}
\usepackage{textcomp}
\usepackage{gensymb}
\usepackage{bm}
\usepackage{braket}
\usepackage{times}
\usepackage[%
  colorlinks=true,
  urlcolor=blue,
  linkcolor=blue,
  citecolor=blue
]{hyperref}

\graphicspath{{./images/}}

\newcommand\imglabelsize{10} 

\begin{document}

\title{\textit{Ab initio} study of the nonlinear optical properties and d.c. photocurrent of the Weyl semimetal \texorpdfstring{TaIrTe$_4$}{TaIrTe4}}

\author{{\'A}lvaro R. Puente-Uriona}
\affiliation{Centro de F{\'i}sica de Materiales, Universidad del Pa{\'i}s Vasco (UPV/EHU), 20018 San Sebasti{\'a}n, Spain}
\author{Stepan S. Tsirkin}
\author{Ivo Souza}
\author{Julen Iba\~{n}ez-Azpiroz}
\affiliation{Centro de F{\'i}sica de Materiales, Universidad del Pa{\'i}s Vasco (UPV/EHU), 20018 San Sebasti{\'a}n, Spain}
\affiliation{Ikerbasque Foundation, 48013 Bilbao, Spain}


\begin{abstract}
We present a first principles theoretical study employing nonlinear response theory to investigate the d.c. photocurrent generated by linearly polarized light in the type-II Weyl semimetal TaIrTe$_4$. We report the low energy spectrum of several nonlinear optical effects. At second-order, we consider the shift and injection currents. Assuming the presence of a built-in static electric field, at third-order we study the current-induced shift and injection currents, as well as the jerk current. We discuss our results in the context of a recent experiment measuring an exceptionally large photoconductivity in this material [J. Ma \textit{et at.}, Nat. Mater. \textbf{18}, 476 (2019)]. According to our results, the jerk current is the most likely origin of the large response. Finally, we propose means to discern the importance of the various mechanisms involved in a time-resolved experiment.
\end{abstract}

\maketitle

\section{Introduction}
Efficient generation of clean energy has become one of the major goals for future sustainability, and renewed attention has been drawn towards the physical processes that transform light into electricity. The bulk photovoltaic effect (BPE) is the nonlinear optical effect that relates the generation of a d.c. photocurrent to light absorption in a homogeneous material. It has its origin in the imbalance of carrier motion along different directions of the Brillouin zone ($\text{BZ}$) \cite{fridkin2001}. At variance with standard methods used for current generation, such as \textit{p-n} junctions, this effect is not tied to the band gap of the material, thus giving rise to large measured photovoltages \cite{osterhoudt2019, spanier2016}.

The quantum mechanical description of nonlinear optical properties is based on response theory, a framework that relates a generated photocurrent density $\bm{j}$ to the illumination field $\bm{E}$ via a photoresponsivity tensor \cite{Butcher1990, aversa1995}. The contributions to the photocurrent are organized in powers of $\bm{E}$, where the response tensors at each order relate the photocurrent contribution to the corresponding power of $\bm{E}$. The BPE is first found at 2$^{\text{nd}}$ order in the illumination field, where the shift~\cite{ibanez-azpiroz2018, vonbaltz1981} and injection \cite{aversa1995} currents are the most relevant contributions. Further orders in perturbation theory can also contribute to the BPE. Recently, attention has been drawn towards 3$^{\text{rd}}$-order effects, which have led to the discovery of current generation mechanisms such as the jerk current, a novel photocurrent effect arising from the divergence of the third-order susceptibility that grows quadratically with illumination time~\cite{fregoso2018}. While the shift current is intuitively understood as an effect that arises from a charge shift accompanying an interband photoexcitation, the jerk current may be interpreted as arising from interband transitions together with the d.c. field accelerating the Bloch electrons.

\begin{figure}
\centering
\includegraphics[width=\columnwidth]{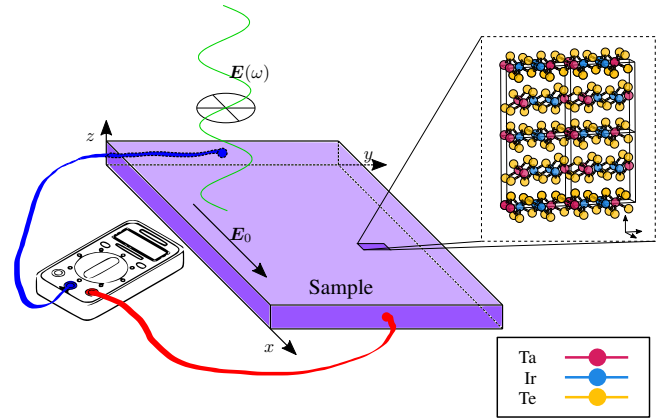}
\caption{Sketch of the measurement setup and unit cell of TaIrTe$_4$ with the $a$, $b$, and $c$ axes along the $x$, $y$, and $z$ axis of the main figure, respectively. We show a sample of TaIrTe$_4$ illuminated by a $xy$ linearly polarized laser. The depicted $\bm{E}_0$ is a built-in electric field whose origin and implications are discussed in Sec. \ref{sec:third_order}.}
\label{fig:exp_drawing}
\end{figure}

The progress in materials synthesis undergone in the past years \cite{phelan2012, Angelsky2022, zhou2021} has allowed measuring the BPE in new structures, like nanotubes \cite{zhang2019}, distorted semiconductors~\cite{yangFlexophotovoltaic2018}, surfaces~\cite{arzate2014, cabellos2011} or Weyl semimetals, which exhibit remarkable optical properties \cite{shao2021, ma2019}. One of the most notable examples corresponds to the type-I Weyl semimetal TaAs, where the quadratic BPE ranks among the largest ever measured \cite{osterhoudt2019}. An exceptionally large BPE has also been recently measured in the type-II Weyl semimetal TaIrTe$_4$ \cite{ma2019}. But unlike in TaAs, the BPE in TaIrTe$_4$ cannot be explained as a 2$^\text{nd}$ order effect due to symmetry arguments. In the experimental setup reported in Ref. \cite{ma2019} the current was measured in the $xy$ plane, as sketched in Fig.~\ref{fig:exp_drawing}. Due to point-group selection rules, the 2$^\text{nd}$ order photoresponsivity tensor vanishes for the in-plane current for a $xy$ linearly polarized laser (this will be further discussed in Sec. \ref{sec:second_order}). As a consequence, the measured photocurrent was attributed to a 3$^\text{rd}$ order BPE.

In this work, we analyze the photocurrent response of TaIrTe$_4$ using first principles calculations. We combine density functional theory (DFT) \cite{kohn1965, giannozziQE2009} with the so-called Wannier interpolation scheme \cite{marzari1997, marzari2012}, which allows us to calculate the photoresponsivity tensors precisely and efficiently. Our analysis is conducted up to 3$^\text{rd}$ order; in addition to the current-induced shift and injection currents considered in Ref. \cite{ma2019}, we study in detail the aforementioned jerk current. As our main result, we find that the jerk current can account for the reported photocurrent data of Ref.~\cite{ma2019} under reasonable experimental conditions. Moreover, we propose an experimental means of verifying the mechanism at play by ultrafast terahertz measurements.

The paper is organized as follows. In Sec. \ref{sec:methods} we summarize the methodology used in our calculations and provide the calculation details. In Sec. \ref{sec:material} we review the electronic structure of TaIrTe$_4$, presenting the energy bands and the position of Weyl points (WPs). In Sec. \ref{sec:second_order} we present our results for the second-order contribution to the BPE, given by the shift and injection currents. In Sec. \ref{sec:third_order} we introduce three possible origins for the large measured photocurrent: the current-induced shift and injection currents, and the jerk current. In Sec. \ref{sec:comp_to_exp_data} we further report a comparison between the current generated by each of the considered mechanisms, and the experimentally measured photocurrent. Lastly, in Sec. \ref{sec:conclusions} we present a summary and the main conclusions of this work.

\section{Methods}\label{sec:methods}
In this section we describe the steps used in our calculations. Our first step consisted of a density functional theory calculation of TaIrTe$_4$ using the package QUANTUM ESPRESSO \cite{giannozziQE2009} to obtain the ground state of the system using a $9\times 3 \times 3$ self consistent field \textit{k}-mesh. Later, we calculated the energy eigenvalues and Bloch eigenfunctions on a $18\times 6 \times 6$ non-self consistent field mesh in the $\text{BZ}$. 
The core-valence interaction was treated by means of fully relativistic projector-augmented-wave pseudopotentials that had been generated with the Perdew-Burke-Ernzerhof exchange-correlation functional. The pseudopotentials were taken from PS library, generated using the ``atomic'' code by A. Dal Corso  v.5.0.2 svn rev. 9415, and the energy cutoff for the plane-wave basis expansion was set at 70 Ry.

In the next step we constructed maximally localized Wannier functions using the WANNIER90 package \cite{pizzi2020}. After discarding the 224 lowest lying bands, we considered a set of 176 disentangled Wannier functions around the Fermi level to span the ``inner" energy range \cite{souza2001} $[-5.2,2.8]\;\text{eV}$. As initial projections, we considered $l=2$ atom-centered orbitals for Ta and Ir atoms, and $l=1$ orbitals for Te atoms. 

To compute the shift, injection and jerk current spectra from Eqs. \eqref{eq:shift_tensor_R}, \eqref{eq:inj_tensor_full} and \eqref{eq:jerk_tensor} below, we employed Wannier interpolation to calculate the transition matrix elements and considered a dense $300\times 100 \times 100$ mesh in the $\text{BZ}$ together with the adaptive broadening scheme, described in Ref. \cite{yates2007}, to handle the Dirac delta functions.

To calculate the jerk current response tensor, we implemented the needed transition matrix elements within the WANNIER90 package. In particular, we followed the prescription of Ref. \cite{yates2007} to evaluate the inverse effective mass in the Wannier basis. To avoid numerical problems near degeneracy points, we regularized energy denominators of the non-abelian part of the Berry connection \textit{gauge} transformation (see Ref. \cite{yates2007}, Eq. (32)) according to
\begin{equation}
\frac{1}{\varepsilon_n-\varepsilon_m}\longrightarrow \frac{\varepsilon_n-\varepsilon_m}{(\varepsilon_n-\varepsilon_m)^2 + \eta^2},
\end{equation}
and we chose $\eta = 40\;\text{meV}$.

\section{Crystal, band structure and topological properties}\label{sec:material}
\begin{figure}
\centering
\includegraphics[width=\columnwidth]{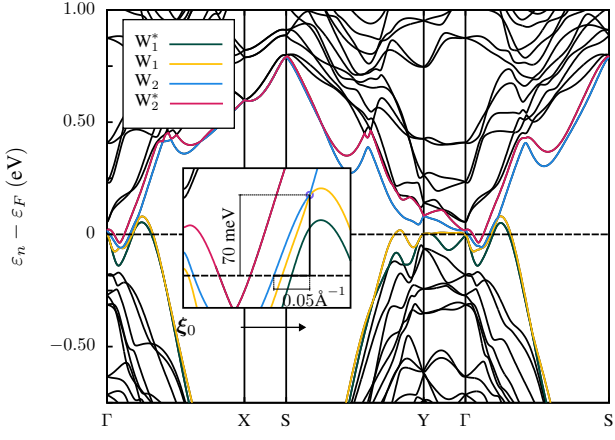}
\caption{Band structure of TaIrTe$_4$ along the primitive orthorhombic fundamental path \cite{setyawan2010} on the $k_z=0$ plane. The four bands crossing the Fermi level are highlighted. We note that the energy separation of the bands is much larger in the $\Gamma-\rm{X}$ or $\rm{X}-\rm{S}$ path than in the $\Gamma-\rm{Y}$ path. The inset shows the band structure along the $(k_x, k_y=0.058,\;k_z=0)\;\text{\r{A}}^{-1}$ line, where two of the four WPs lying on the $k_z=0$ plane are located. The symbol $\bm{\xi}_0$ refers to the $(k_x = 0, k_y=0.058,\;k_z=0)\;\text{\r{A}}^{-1}$ point, and the arrow denotes the positive $k_x$ direction.}
\label{fig:band_structure}
\end{figure}

TaIrTe$_4$ is a thoroughly studied type-II Weyl semimetal~\cite{zhouWeyl2018}, with space group Pmn2$_1$ (No. 31, noncentrosymmetric). It is formed by a layered structure with octahedrally coordinated Ir and Ta atoms, and the orthorhombic lattice has experimental constants $a = 3.80\;\text{\r{A}}$, $b = 12.47\;\text{\r{A}}$, $c = 13.24\;\text{\r{A}}$. Fig. \ref{fig:band_structure} shows the interpolated band structure in the neighborhood of the Fermi level $\varepsilon_F$. Inspection of the figure reveals metallic character along the $\Gamma-\rm{Y}$ path while describing insulating character along those points far away form $\Gamma$ in the $\Gamma-\rm{X}$ and $\Gamma-\rm{S}$ paths. We attach a comparison between the DFT energy eigenvalues and those obtained in the Wannierization procedure in Appendix \ref{sec:appendix} at Fig. \ref{fig:scf_vs_wann}.

The space group Pmn2$_1$ has 4 different symmetry operations, $\mathds{1},\;M_x,\; \{C_{2z}|(1/2,0,1/2)\}$ and $\{M_y|(1/2,0,1/2)\}$ \cite{itfc_sgs}. The point-group symmetry operations associated with these space-group operations form a little group when combining with time-reversal symmetry and allow degeneracies to occur at the symmetry-invariant planes. These degeneracies give rise to WPs in the band structure.

We find that TaIrTe$_4$ hosts a total of 12 WPs close to the $\Gamma$ point and near $\varepsilon_F$, four of which are in the $k_z=0$ plane at $\text{WP}_1=\{(\pm 0.189, \pm 0.058, 0)\}\;\text{\r{A}}^{-1}$, and the other eight at $\text{WP}_2=\{(\pm 0.079, \pm 0.021, \pm 0.053)\}\;\text{\r{A}}^{-1}$, in good agreement with a previous calculation \cite{zhouWeyl2018}. The first four WPs lie $\approx 70\;\text{meV}$ above $\varepsilon_F$, while the other eight lie $\approx 70\;\text{meV}$ below $\varepsilon_F$. In Fig. \ref{fig:band_structure}, we show the band structure of the crystal along high symmetry-lines on the $k_z=0$ plane, highlighting the bands that give rise to the WPs (labelled W$_{1}$, W$_{2}$) and their spin-doubled bands (labelled W$_{1}^*$, W$_{2}^*$). 

\section{Second-order response}
\label{sec:second_order}
In this section we analyze the 2$^{\text{nd}}$ order mechanisms that generate a BPE. We take into account the well-known shift- and injection-current mechanisms.

\subsection{Shift current}

The shift current corresponds to the intrinsic interband d.c. part of the 2$^\text{nd}$ order BPE \cite{sipe2000}. Its photocurrent density is expressed as
\begin{equation}\label{eq:shift_photocurrent}
j^a_{\text{shift}} = 2\sum_{bc}\sigma^{abc}(\omega)\text{Re}\left[E^b(\omega)E^c(-\omega)  \right],
\end{equation}
where $a$, $b$, and $c$ are Cartesian indices and $\sigma^{abc}$ is the shift current photoconductivity tensor, which is symmetric under the $b\Leftrightarrow c$ exchange and transforms like the piezoelectric tensor \cite{Nye1984}. It is given by \cite{sipe2000,ibanez-azpiroz2018}
\begin{equation}\label{eq:shift_tensor_R}
\begin{split}
\sigma^{abc}(\omega)& = -\frac{i\pi e^3}{4\hbar^2}\int_{\text{BZ}}\frac{d^3 \bm{k}}{(2\pi)^3}\sum_{nm}f_{nm} \big[ r_{mn}^b r_{nm;a}^c \\
& + (b\Leftrightarrow c) \big] \times \big[\delta(\omega_{mn}+\omega)+\delta(\omega_{mn}-\omega)\big].
\end{split}
\end{equation}
Here $f_{nm} = f_n - f_m$ is the difference between the Fermi occupation factors and $\omega_{nm} = \left(\varepsilon_n - \varepsilon_m\right)/\hbar$ (the optical excitation frequency). The other relevant quantities appearing in the integral involve the dipole matrix elements and its ``generalized derivative". These are given by
\begin{align}
r_{nm}^a & = (1-\delta_{nm})A_{nm}^a, \\
r_{nm;b}^a & = \partial_b r_{nm}^a -i\left(A_{nn}^b - A_{mm}^b \right)r_{nm}^a
\end{align}
respectively, where $A_{nm}^a$ is the Berry connection,
\begin{equation}
A_{nm}^a = i\braket{n|\partial_a|m}.
\end{equation}
In the equations above we have omitted the $\bm{k}$ dependence and $\partial_a$ stands for $\partial/\partial k^a$, and $\ket{n}, \ket{m}$ denote the cell-periodic Bloch states.

\begin{figure}
\centering
    \includegraphics[width=\columnwidth]{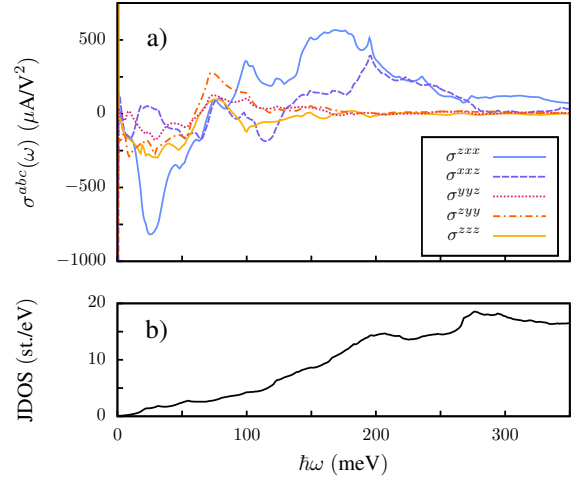}
\caption{a) Spectra of all the symmetry-allowed components of the shift current tensor for TaIrTe$_4$. b) Joint Density of States (JDOS).}
\label{fig:sc_all}
\end{figure}

Due to the orthorhombic crystal structure, only a few of the tensor components are symmetry-allowed and independent \cite{Boyd2008}. These are $xxz, yyz, zxx, zyy, zzz$; therefore, we note that $\sigma^{abc}$ vanishes if the indices $a,b,c=\{x,y\}$. The nonzero tensor component spectra and the joint density of states (JDOS) are shown in Fig. \ref{fig:sc_all}. Inspection of the figure reveals that $\sigma^{zxx}$ is the dominant component in the $[0,350]\;\text{meV}$ energy region, reaching up to $500\;\mu\text{A/V}^2$ at $\approx 200\;\text{meV}$, which also corresponds to a maximum of the JDOS. We highlight that these spectra are an order of magnitude larger than the dominant response of most materials, such as GaAs~\cite{ibanez-azpiroz2018} or BaTiO$_3$ \cite{young2012}, and comparable to the dominant response in the type-I Weyl semimetal TaAs~\cite{osterhoudt2019} and of single layer monochalcogenides Ge and Sn \cite{rangel2017}.

\begin{figure}
\centering
\includegraphics[width=\columnwidth]{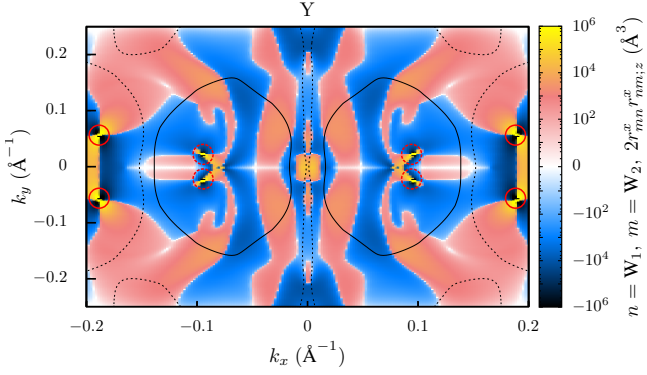}
\caption{Matrix elements $2 r_{mn}^x r_{nm;z}^x $ in the $k_z=0$ plane for the bands $n=\text{W}_1$ and $m=\text{W}_2$. The Fermi contours for the $\text{W}_1$ and $\text{W}_2$ bands are highlighted along the plane with dotted and solid lines, respectively. The regions where the $\text{W}_1$ and $\text{W}_2$ bands are respectively  occupied and unnocupied fall in between dashed and solid Fermi cuts. The largest values \big($> 10^5\;\text{\r{A}}^3$\big) appear in the neighborhood of the WPs, which are marked with solid (WP$_1$) or dashed (WP$_2$) red circles.}
\label{fig:sc_zxx_integrand}
\end{figure}

It has been previously shown \cite{ibanez-azpiroz2018, ma2021, cook2017} that the shift current can be strongly enhanced by topologically-protected band crossings, such as Weyl nodes. To check if this is the case for TaIrTe$_4$ as well, in Fig. \ref{fig:sc_zxx_integrand} we display the matrix element $2 r_{mn}^x r_{nm;z}^x$ for $n=\text{W}_1$, $m=\text{W}_2$ transition across the $k_z=0$ plane. The heat map shows clear peaks in the neighborhood of the WPs (marked with red circles). Closer inspection reveals secondary peaks at the $(k_x,k_y)$ projections of the other eight WPs (marked with dashed red circles) at $k_z\neq 0$. But despite their large magnitude, direct transitions to and from the WPs do not contribute to the shift current due to the occupation factors, as can be deduced from the Fermi cuts in Fig.~\ref{fig:sc_zxx_integrand}. The net shift current at low energies is instead determined by other regions in the $\text{BZ}$ where transitions between metallic bands are allowed.

\subsection{Injection current}
\label{sec:inj_curr}

The injection current corresponds to the rate of change of the intrinsic interband d.c. part of the 2$^\text{nd}$ order BPE \cite{sipe2000}. The photocurrent density reads
\begin{equation}\label{eq:inj_photocurrent}
j^a_{\text{inj}} = \tau\sum_{bc}\eta^{abc}(\omega)E^b(\omega)E^c(-\omega).
\end{equation}
In writing Eq. \eqref{eq:inj_photocurrent}, we have assumed that the relaxation time is characterized by a lifetime $\tau$, which is expected to be of the order of momentum relaxation time, $\tau = 10^{-14}\;\text{s}$ \cite{ma2019}. $\eta^{abc}(\omega)$ in Eq. \eqref{eq:inj_photocurrent} is the injection current photoconductivity tensor, which has no definite parity under the $b\Leftrightarrow c$ exchange. It is given by \cite{aversa1995}
\begin{equation}\label{eq:inj_tensor_full}
\begin{split}
\eta^{abc}(\omega)& = -\frac{\pi e^3}{\hbar^2}\int_{\text{BZ}}\frac{d^3 \bm{k}}{(2\pi)^3}\sum_{nm}f_{nm} \frac{\partial \omega_{mn}}{\partial k^a} r_{nm}^b r_{mn}^c \\
&\times \delta(\omega_{mn}-\omega).
\end{split}
\end{equation}

By virtue of Eq. \eqref{eq:inj_photocurrent}, the injection current mechanism can generate a photocurrent regardless of the light polarization. We proceed as in Ref. \cite{wang2020} and separate $\eta^{abc}$ into symmetric and antisymmetric components under the $b\Leftrightarrow c$ exchange,
\begin{subequations}
\label{eq:inj_sym_asym_components}
\begin{align}
\eta^{abc}_S(\omega) & = \frac{1}{2}\left[\eta^{abc}(\omega) + \eta^{acb}(\omega)\right],\label{eq:inj_sym}\\
\eta^{abc}_A(\omega) & = \frac{1}{2}\left[\eta^{abc}(\omega) - \eta^{acb}(\omega)\right].
\label{eq:inj_asym}
\end{align}
\end{subequations}
The tensors $\eta^{abc}_A$, $\eta^{abc}_S$  are purely imaginary and real, respectively. The antisymmetric (A) part describes a current that is equal and opposite for left‐ and right- circularly polarized light, and hence it vanishes for linearly polarized light. As for the symmetric (S) part, it generates equal currents for left‐ and right‐circularly polarized light. 

\begin{figure}
\centering
    \includegraphics[width=\columnwidth]{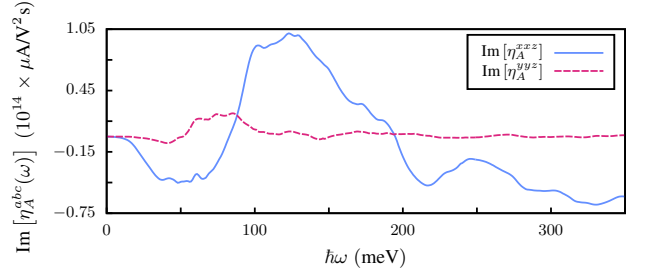}
\caption{Spectra of the two symmetry-allowed independent components of the antisymmetric part of the injection current tensor for TaIrTe$_4$.}
\label{fig:icA_all}
\end{figure}

Due to the crystal structure, only the $xxz$ and $yyz$ components of $\eta^{abc}_A$ are nonzero and independent. The spectra are shown in Fig. \ref{fig:icA_all} and are similar in magnitude to the dominant response of other bulk materials, such as CdS and CdSe \cite{nastos2010}, but more than an order of magnitude smaller than type-I Weyl semimetals RhSi and PtAl \cite{le2020}. We recall that the contribution to the photocurrent due to $\eta^{abc}_A$ vanishes under application of linearly polarized light and as such is not directly relevant for comparison with the experiment of Ref. \cite{ma2019}. In the case of $\eta^{abc}_S$, \textit{all} the components vanish given that it is odd under the 
time reversal symmetry operation \cite{zhangSwitchable2019}. 

\section{Third-order responses}
\label{sec:third_order}

Owing to the point-group symmetry of the TaIrTe$_4$ crystal, the shift and injection currents studied in the previous section feature nonzero tensor components only if at least one of the tensor indices $a, b, c$ equals $z$. In the experiment reported in Ref. \cite{ma2019}, however, current could only be measured for the  ``in-plane"  components $j^x$ and $j^y$ under light polarized in the $xy$ plane. Therefore, as noted in Ref. \cite{ma2019}, the shift and injection current mechanisms are unable to account for the measured ``in-plane" photocurrent, which must have a different origin. 

In this section we study three particular contributions to the third-order BPE: the jerk current, and the current-induced (CI) shift and injection currents. These effects require the presence of a $\bm{E}_0$ d.c. field in the material, and feature symmetry-allowed tensor components that can potentially account for the measured ``in-plane" photocurrent \cite{ma2019}. The CI shift and injection current mechanisms were proposed in Ref. \cite{ma2019} (although not named as such), and arise from carrier imbalance generated by $\bm{E}_0$ in the shift and injection currents. On the other hand, the jerk current originates from the combined effects of optical pumping by a laser field $\bm{E}(\omega)$, and the d.c. field $\bm{E}_0$ accelerating Bloch electrons.

The presence of the static field is argued in Ref. \cite{ma2019} to be a consequence of the experimental configuration employed, having a possible origin in the workfunction difference between the sample and the metal contacts used for measurement, as we illustrate in Fig. \ref{fig:exp_drawing}. While the magnitude of $E_0$ cannot be experimentally controlled, the expected realistic values lie within the $E_0 \in [10^4,10^6 ]\;\text{V/m}$ range and the direction is taken to be aligned within the crystallographic $a$ axis \cite{ma2019}.

\subsection{Jerk current}
We begin by considering the jerk current, a 3$^\text{rd}$ order response recently described in Refs. \cite{fregoso2018, ventura2021, mendoza2020}. This d.c. current originates from the action of an optical field in combination with a d.c. bias. The induced current density is
\begin{equation}\label{eq:jerk_photocurrent}
j^a_{\text{jerk}} = \tau^2\sum_{bcd}g_{\text{jerk}}^{abcd}(\omega)E^b(\omega)E^c(-\omega)E^d(0),
\end{equation}
with the jerk current tensor $g_{\text{jerk}}^{abcd}(\omega)$ \cite{ventura2021,mendoza2020}
\begin{equation}\label{eq:jerk_tensor}
\begin{split}
g_{\text{jerk}}^{abcd}(\omega) &= \frac{2\pi e^4}{\hbar^3}\int_{\text{BZ}}\frac{d^3 \bm{k}}{(2\pi)^3}\sum_{nm}f_{nm} \frac{\partial^2 \omega_{nm}}{\partial k^a \partial k^d} r_{nm}^b r_{mn}^c \\ &\times\delta(\omega_{nm}-\omega).
\end{split}
\end{equation}
The 4$^\text{th}$ order response tensor in Eq. \eqref{eq:jerk_tensor} is symmetric under $a\Leftrightarrow d$ and $b\Leftrightarrow c$ exchanges, and it can be interpreted using a simple phenomenological model \cite{mendoza2020, ventura2021}: the $\partial^2_{ad}\omega_{nm} E_0^d$ term corresponds to carrier acceleration along the $\text{BZ}$ which involves the difference between the inverse effective masses of the bands $n, m$, while $f_{nm}r_{nm}^b r_{mn}^c \delta(\omega_{mn}-\omega)E^b(\omega)E^c(-\omega)$ corresponds to carrier pumping from a filled to an empty band.

\begin{figure}
\centering
\includegraphics[width=\columnwidth]{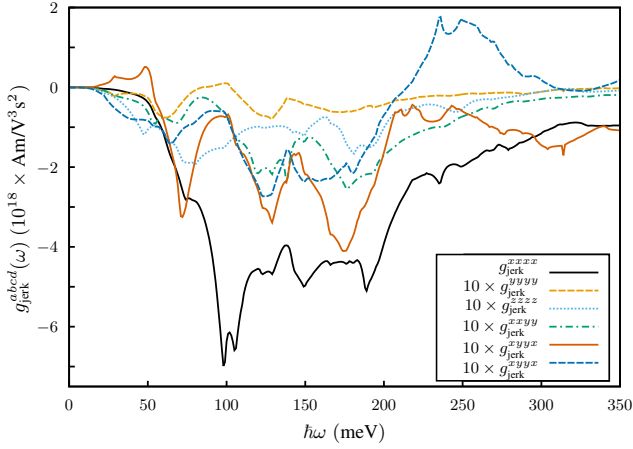}
\caption{Spectra of the ``main-axis" and the ``in-plane" elements of the jerk current tensor for TaIrTe$_4$.}
\label{fig:jerk_spr}
\end{figure}

Due to point-group arguments, only those components with indices all equal, or equal in pairs are symmetry-allowed~\cite{itfc_sgs}. In Fig. \ref{fig:jerk_spr} we show the ``main axis" elements (those with all indices equal) and the symmetry-allowed elements that partake in the generation of ``in-plane" current \big(those with $a,b,c,d = \{x,y\}$\big). In Fig. \ref{fig:jerk_other} at Appendix \ref{sec:appendix} we show all the other nonzero components. We highlight that the $g_{\text{jerk}}^{xxxx}$ ``main axis" spectrum is the dominant one, being an order of magnitude larger than the rest, and three orders of magnitude larger than recently studied materials such as GeS~\cite{mendoza2020}.

\begin{figure}
\centering
\includegraphics[width=\columnwidth]{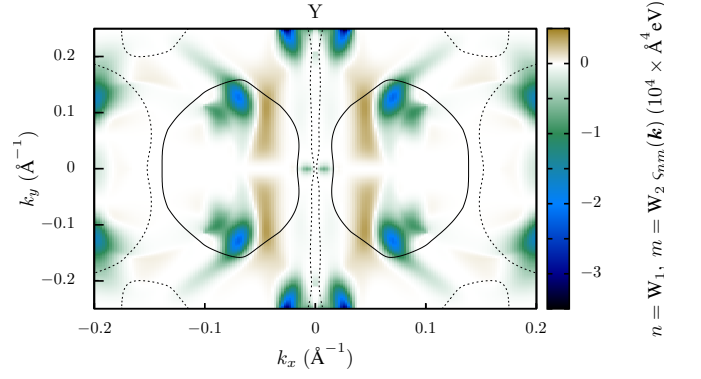}
\caption{Matrix elements $\varsigma_{nm}(\bm{k}) = |r_{nm}^x|^2\;\varepsilon_{nm,xx}$ appearing in the integrand of Eq. \eqref{eq:jerk_tensor} corresponding to the $n=\text{W}_1\leftrightarrow m=\text{W}_2$ transition for the $g_{\text{jerk}}^{xxxx}$ component in the $k_z=0$ plane. The Fermi contours for the bands $\text{W}_1,\;\text{W}_2$ are highlighted along the plane with dotted and solid lines, respectively. The regions where the $\text{W}_1$ and $\text{W}_2$ bands are respectively  occupied and unnocupied fall in between dashed and solid Fermi cuts. Note that there are no special contributions from the WPs and that the hotspots near the $k_x=0$ line are responsible for major contributions. The inset shows the shape of the bands responsible for this transition in the proximity of the hotspots.}
\label{fig:jerk_integrand_wb}
\end{figure}

We have verified that approximately half of the contribution to the jerk current in the low-energy range has its origin in transitions between the Weyl bands. The corresponding matrix element for the $n=\text{W}_1\leftrightarrow m=\text{W}_2$ transition in the $k_z=0$ plane is shown in Fig. \ref{fig:jerk_integrand_wb}. The hotspots close to the $\rm{Y}$ point dominate the heatmap plot. The origin of these hotspots can be traced back to a constructive interference between the dipole matrix element $|r_{nm}^x|^2$ and the inverse effective mass term $\varepsilon_{nm,xx}$, which takes place in a region where $\varepsilon_{\text{W}_{2}}-\varepsilon_{\text{W}_{1}}\simeq 100$~meV. This electronic structure effect is the main responsible for the major peak of the ``in-plane" $g_{\text{jerk}}^{xxxx}$ spectrum in Fig. \ref{fig:jerk_spr}.

\subsection{Comparison with current-induced contributions}
\label{sec:contrib_ci}
In this section we consider again the shift and injection current mechanisms, and we generalize them to account for the presence of a d.c. electric field. Such a field allows electron transport across the $\text{BZ}$, breaking the equivalence between $\bm{k}$ and $-\bm{k}$ (time-reversal symmetry breaking) and thus shifting the Fermi occupation factors according to $f_n(\bm{k})\longrightarrow f_n(\bm{k} + \bm{k}_0)$. The \textit{k}-shift is given by the semiclassical prescription
\begin{equation}\label{eq:kshift}
\bm{k}_0 = -\frac{e}{\hbar}\tau \bm{E}_0.
\end{equation}
If the magnitude of the shift is small enough when compared with the dimension of the $\text{BZ}$, the Fermi occupation factors may be expanded in a Taylor series. Keeping terms up to linear order in $\bm{E}_0$, 
\begin{equation}\label{eq:taylor_occupations}
\begin{split}
f_n(\bm{k} + \bm{k}_0) & \approx f_n(\bm{k}) + \sum_i k_0^i \cdot \frac{\partial f_n(\bm{k})}{\partial k^i} \\
& =  f_n(\bm{k}) + \sum_i \hbar k_0^i v^i_{n}\delta(\varepsilon_n-\varepsilon_F),
\end{split}
\end{equation}
with $v^i_{n} = \partial_i \omega_n$ the band velocity. The first term corresponds to the equilibrium occupation, while the second one, the (linear) CI term, relates linearly to the static field $\bm{E}_0$. Note that the delta function is to be understood as a Gaussian distribution centered at the Fermi energy with a thermal smearing of $25\;\text{meV}$ \cite{ma2019}. Previous works \cite{mendoza2021} have employed a similar method to calculate the contribution to the photocurrent of the bands that cross the Fermi surface.

In the case of the CI shift current, the photocurrent density is obtained by replacing the equilibrium occupation factors $f_n$ 
in Eq.~\eqref{eq:shift_tensor_R}
by the second term in the right hand side of Eq. \eqref{eq:taylor_occupations}, leading to 
\begin{equation}\label{eq:shift_LCI_photocurrent}
j^a_{\text{CI sc}} = 2\tau\sum_{bcd}g^{abcd}_{\text{sc}}(\omega)\text{Re}\left[E^b(\omega)E^c(-\omega)  \right]E^d(0),
\end{equation}
with $g^{abcd}_{\text{sc}}$ given by
\begin{equation}\label{eq:shift_LCI_tensor}
\begin{split}
g^{abcd}_{\text{sc}}(\omega)& = \frac{i\pi e^4}{4\hbar^2}\int_{\text{BZ}}\frac{d^3 \bm{k}}{(2\pi)^3}\sum_{nm} \big[ r_{mn}^b r_{nm;a}^c + (b\Leftrightarrow c) \big]\\
& \times \big[v^d_{n}\delta(\varepsilon_n-\varepsilon_F)-v^d_{m}\delta(\varepsilon_m-\varepsilon_F)\big]\\
& \times \big[\delta(\omega_{mn}+\omega)+\delta(\omega_{mn}-\omega)\big].
\end{split}
\end{equation}
We note that $j^a_{\text{CI shift}}$ grows linearly with the momentum relaxation time.

To investigate the linear CI injection current, we consider $\eta^{abc}_S$ given in Eq. \eqref{eq:inj_sym}, which is activated by the broken time-reversal symmetry under the d.c. bias $\bm{E}_0$ and, unlike $\eta_A^{abc}$ in Eq. \eqref{eq:inj_asym} it is capable of generating a photocurrent under linearly polarized light. The corresponding photocurrent density is expressed as
\begin{equation}\label{eq:inj_LCI_photocurrent}
j^a_{\text{CI ic}} = 2\tau^2\sum_{bcd}g^{abcd}_{\text{ic}}(\omega)\text{Re}\left[E^b(\omega)E^c(-\omega)  \right]E^d(0),
\end{equation}
with $g^{abcd}_{\text{ic}}$ given by 
\begin{equation}\label{eq:inj_LCI_tensor}
\begin{split}
g^{abcd}_{\text{ic}}(\omega)& = \frac{\pi e^4}{2\hbar^2}\int_{\text{BZ}}\frac{d^3 \bm{k}}{(2\pi)^3}\sum_{nm}\frac{\partial \omega_{mn}}{\partial k^a} \big[r_{nm}^b r_{mn}^c + (b\Leftrightarrow c) \big] \\
& \times \big[v^d_{n}\delta(\varepsilon_n-\varepsilon_F)-v^d_{m}\delta(\varepsilon_m-\varepsilon_F)\big]\\
& \times \delta(\omega_{mn}-\omega),
\end{split}
\end{equation}
which is obtained by replacing $f_n$ in Eq.~\eqref{eq:inj_tensor_full} by the
second term in Eq.~\eqref{eq:taylor_occupations}.
At variance with the shift contribution, $j^a_{\text{CI inj.}}$ in Eq. \eqref{eq:inj_LCI_photocurrent} grows quadratically with momentum relaxation time and $g_{\text{ic}}$ in Eq. \eqref{eq:inj_LCI_tensor} has the same units as the jerk current tensor $g_{\text{jerk}}$.

\begin{figure}
\centering
\includegraphics[width=\columnwidth]{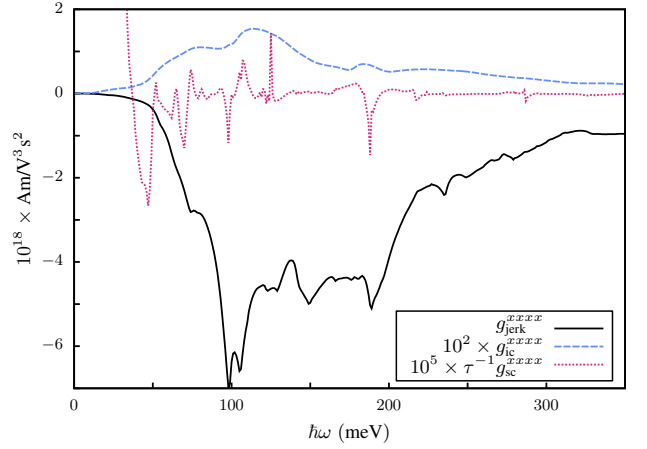}
\caption{Comparison between the tensor components $g_{\text{jerk}}^{xxxx}$, $g^{xxxx}_{{\text{ic}}}$ and $\tau^{-1} g^{xxxx}_{{\text{sc}}}$, corresponding to jerk, CI injection and CI shift current generation mechanisms, respectively.}
\label{fig:spectra_comparison}
\end{figure}

Both tensors $g^{abcd}_{{\text{sc}},\text{ic}}$ feature symmetry-allowed ``in-plane" components \cite{itfc_sgs} which may generate  ``in-plane" current under the action of light polarized in the $xy$ plane. In order to investigate the magnitude of their current, we focus on the dominant components $\tau^{-1} g^{xxxx}_{{\text{sc}}}$, $g^{xxxx}_{{\text{ic}}}$ and compare them to the jerk current contribution $g_{\text{jerk}}^{xxxx}$ in Fig. \ref{fig:spectra_comparison}.

The figure shows that the jerk current is larger than the CI injection (shift) current by a factor of at least $10^2$ ($10^5$) in the $[50, 200]\;\text{meV}$ energy range. For energies above $200\;\text{meV}$, the CI contributions become at least an order of magnitude smaller. We note that the divergence of $\tau^{-1} g^{xxxx}_{{\text{sc}}}$ as $\omega \longrightarrow 0 $ is a numerical artifact caused by the implementation of the ``generalized derivative" $r^a_{nm;b}$, and it tends to disappear for increasing \textit{k}-point grids. Therefore, according to our results, the jerk current dominates over the CI shift and injection currents in the whole low energy range.

\subsection{Angular dependence of the photocurrent}\label{sec:comp_to_exp_data}
Given the finite magnitude of the ``in-plane" jerk, CI injection and CI shift-current components, we now proceed to estimate the photocurrent generated by these mechanisms under the experimental conditions of Ref. \cite{ma2019}. In that work, linearly polarized light in the $xy$ plane was employed, $\bm{E} = E(\omega)\left(\cos\theta\; \hat{x}+\sin\theta\; \hat{y} \right) + \text{h.c}$.

The total current generated by any of the above-mentioned mechanisms can be generically written as
\begin{equation}
\label{eq:current_components}
\begin{alignedat}{2}
J_{\text{mec}}^a(\theta,\omega) &=E_0 C &\tau^{n} &g_{\text{mec}}^{axxx}(\omega) \cos^2\theta\\
&+ 2E_0 C &\tau^{n} &g_{\text{mec}}^{axyx}(\omega) \cos\theta\sin\theta\\
&+ E_0 C &\tau^{n} &g_{\text{mec}}^{ayyx}(\omega) \sin^2\theta,
\end{alignedat}
\end{equation}
where $\text{mec} = \text{jerk, ic}$ or $\text{sc}$. In writing Eq. \eqref{eq:current_components} we have exploited the $b \Leftrightarrow c$ symmetry of the tensors, and the exponent $n$ contains the time dependence of the absorption process of the considered mechanism: $n = 1$ for $g_{\text{mec}} = g_{\text{sc}}$, and $n = 2$ for $g_{\text{mec}} = g_{\text{jerk}},\;g_{\text{ic}}$. The constant $C$ is given by
\begin{equation}\label{eq:amplitude_to_power}
C = S_{\text{sec}}|E(\omega)|^2  =  \frac{S_{\text{sec}}}{S_{\text{spot}}}\frac{2P}{c\varepsilon_0}.
\end{equation}
In addition to the speed of light $c$ and the vacuum permittivity $\varepsilon_0$, $C$ encompasses the characteristics of the experimental setup: $S_{\text{sec}}$ is the cross section of the sample, while $S_{\text{spot}}$ and $P$ are the spot size and the power of the incident laser, respectively.

The experiment of Ref. \cite{ma2019} considered two different power regimes for measurements. In the low-power regime, ranging from $\text{nW}$s to $\mu\text{W}$s, the reported photoresponsivity $\kappa$ \cite{cook2017} was of the order of $\kappa\approx 100\;\text{mAW}^{-1}$, while in the high-power regime, ranging from $\mu\text{W}$s to $\text{mW}$s, the photoresponsivity was of the order of $\kappa\approx 1\;\text{mAW}^{-1}$. Since our approach is based on perturbation theory with respect to the amplitude of the electric field, we focus on understanding the low-power regime.

We use the experimental value for the laser power $P = 480\;\text{nW}$, and estimate the ratio $S_{\text{sec}}/S_{\text{spot}}$ with the aid of Fig.~2b of Ref. \cite{ma2019}. We do this by using $S_{\text{sec}}/S_{\text{spot}} = J/ \kappa P $, which relates the photocurrent to laser power via the photoresponsivity $\kappa$ \cite{cook2017}. Fig.~2b of Ref. \cite{ma2019} reports that for $P\approx 200\;\text{nW}$, $J\approx 5\;\text{nA}$ and $\kappa\approx 130\;\text{mAW}^{-1}$, hence $S_{\text{sec}}/S_{\text{spot}} \approx 0.18$. Using Eqs. \eqref{eq:current_components} and \eqref{eq:amplitude_to_power}, we have calculated the angular polarization dependence of the photocurrent generation mechanisms and compared it with the experimental data. 

\begin{figure}
\centering
\includegraphics[width=\columnwidth]{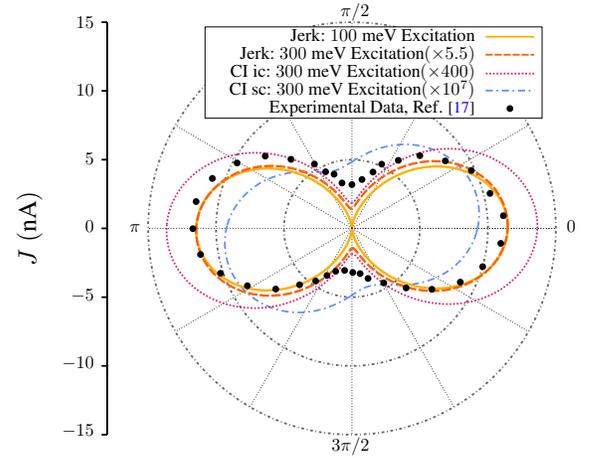}
\caption{Comparison between the experimentally measured photocurrent (black) at $\hbar\omega=300\;\text{meV}$, the jerk photocurrent at $\hbar\omega=100,\;300\;\text{meV}$ (yellow solid, orange dashed) excitations, the CI injection current (magenta) at $\hbar\omega=300\;\text{meV}$ and the CI shift current (blue) at $\hbar\omega=300\;\text{meV}$. We have assumed values for the d.c. electric field and lifetime of $\bm{E}_0 = 2.82\times 10^5 \hat{x}\;\text{V/m}$ and $\tau=10^{-14}\;\text{s}$. A factor of $5.5$, $400$, $10^7$ have been applied to the $300\;\text{meV}$ jerk, current induced (linear) injection and shift excitations, respectively, which may be absorbed in the value of $E_0$.}
\label{fig:polar_photocurrent}
\end{figure}

For the built-in static electric field $\bm{E}_0$, we have considered a moderate value of $\bm{E}_0 = 2.82\times 10^5 \hat{x}\;\text{V/m}$, which lies in the middle of the experimentally expected range $E_0 \in [10^4,10^6 ]\;\text{V/m}$ \cite{ma2019}. We show the comparison between theoretical and experimental values in Fig. \ref{fig:polar_photocurrent} as a function of the polarization angle $\theta$. This is done for two different excitation energies $\hbar\omega=100\;\text{meV}$ and $\hbar\omega=300\;\text{meV}$ in the case of the jerk current, which correspond to the peak of the jerk current (Fig. \ref{fig:jerk_spr}) and the experimentally measured peak \cite{ma2019}, respectively. In addition, we also show the analogous polar plot for the CI injection and shift currents at $\hbar\omega=300\;\text{meV}$, respectively. As expected from Fig. \ref{fig:spectra_comparison}, the jerk current is the dominating contribution to the photocurrent.


The jerk current curve shows an elongated shape around $\theta=0$ for both frequencies. This is due to the large magnitude of $g_{\text{jerk}}^{xxxx}$ as compared to the rest of the ``in-plane" jerk current tensor components (Fig. \ref{fig:jerk_spr}). It is remarkable that under reasonable assumptions for the magnitude of $\bm{E}_0$ and $\tau$, the calculated jerk current matches both the value as well as the angular dependence of the experimentally measured photocurrent, with only small deviations at polarization angles $\theta = \pi/2,3\pi/2$. While the frequency of the peak jerk current differs by $\approx 200\;\text{meV}$ with the experimental one, it is on the expected range of accuracy for a DFT excitation energy. Our results therefore indicate that the jerk current is a likely candidate for the large photoresponsivity measured in Ref.~\cite{ma2019}. In the case of the CI injection-current, the shape agrees nicely with the experimental measurement, but it would require an unreasonably large electric field of $E_0 > 10^7 \;\text{V/m}$ to account for the experimentally measured current magnitude. As for the CI shift-current, the shape is significantly tilted at around $\theta \approx 15\degree$ and the size is orders of magnitude too small.

We conclude with a comment on the  momentum relaxation time $\tau$ and its possible effect on the hierarchy of the analyzed  physical processes. For the above discussion we have considered $\tau=10$ fs,  a typical value for metals assumed also in Ref.~\onlinecite{ma2019}. Previous works have reported a wide range in other materials, from $\simeq 2$ fs in BaTiO$_3$~\cite{dai2021} to hundreds of femtoseconds in CrI$_3$~\cite{zhangSwitchable2019}. We note that, even if we consider $\tau \in [1,10^{3}]\; \text{fs}$ for TaIrTe$_{4}$, the jerk current remains the dominant third-order d.c. contribution in the whole range; the ratio to the CI injection current is unchanged since both are proportional to $\bm{E}_0\tau^2$ (see Eqs.~\ref{eq:jerk_photocurrent} and~\ref{eq:inj_LCI_photocurrent}), while the CI shift current, which is proportional to $\bm{E}_0\tau$ (see Eq.~\ref{eq:shift_LCI_photocurrent}), is $\simeq4$ orders of magnitude smaller than the jerk current even for $\tau =1$ fs. Therefore, our conclusions hold for a wide range of the momentum relaxation time.

\section{Conclusion and Outlook}\label{sec:conclusions}
We have analyzed three different third-order contributions to the d.c. photocurrent of type-II Weyl semimetal TaIrTe$_4$, and discussed them in the context of recent experimental measurements \cite{ma2019}. These mechanisms are the jerk current, CI injection current and CI shift current. The first is a current generation mechanism proposed recently in \cite{fregoso2018, ventura2021, mendoza2020}, and the other two are generalizations, proposed in Ref. \cite{ma2019}, of the well-known injection and shift current mechanisms. All the above-mentioned mechanisms assume the presence of a d.c. field acting on the material. Our method is based on employing \textit{ab inito} calculations to obtain the band structure and the Berry connection of the material. These quantities are later used to calculate the optical response spectra using Wannier interpolation. We have found that out of the considered mechanisms, the jerk current matches both the value as well as the shape of the experimental data under reasonable assumptions for the magnitudes the built-in static electric field $\bm{E}_0$ and of the relaxation time $\tau$. Our calculations show that the large magnitude of the jerk photocurrent is caused by a electronic structure effect, associated to a constructive interference between the dipole matrix element and the difference between the inverse effective masses of Weyl bands.

According to the definition of the jerk current in Eq. \eqref{eq:jerk_photocurrent} and the definition of the CI injection current in Eq. \eqref{eq:inj_LCI_photocurrent}, the generated current grows quadratically with illumination time in both cases. As a characteristic difference between these two mechanisms, the CI injection-current can generate a circular photogalvanic effect  whereas the jerk current cannot. As for the CI shift current, it grows linearly with illumination time and cannot generate a 
circular photogalvanic effect. We propose these differences to discern the origin of the main source of the 3$^{\text{rd}}$-order BPE in TaIrTe$_4$. This could potentially be achieved by using fast laser pulses, which are nowadays capable of measuring the time evolution of the signal the picosecond to femtosecond scale \cite{shallcross2022, bowlan2014, bertoni2016, ishioka2010, wickramaratne2022}.

\section{Acknowledgments}
This project has received funding from the European Union’s Horizon 2020
research and innovation programme under the European Research Council (ERC)
grant agreement No 946629 and from Grant No. PID2021-129035NB-I00 funded by
MCIN/AEI/10.13039/501100011033.

\appendix
\section{Supplementary numerical results}\label{sec:appendix}

\begin{figure}
\centering
\includegraphics[width=\columnwidth]{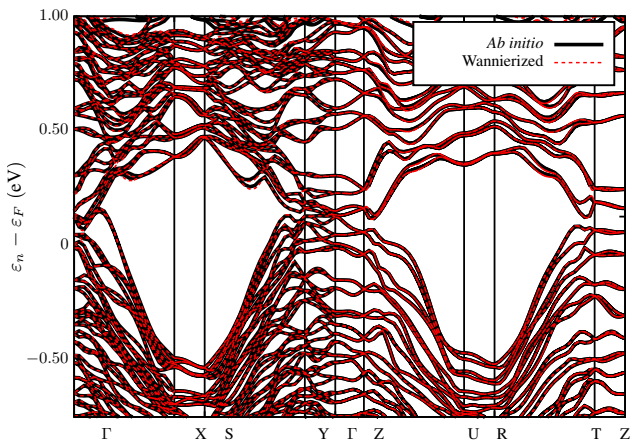}
\caption{Comparison between the band structure obtained using the DFT QUANTUM ESPRESSO package and the band structure obtained from post processing the DFT results using the WANNIER90 package for TaIrTe$_4$. We show the bands in the $[-0.75,1.00]\;\text{eV}$ range, which accounts for most of theenergy range where optical transitions take place in the considered $[-5.2,2.8]\;\text{eV}$ ``inner" energy range. The bands are shown along the primitive orthorhombic fundamental path \cite{setyawan2010}.}
\label{fig:scf_vs_wann}
\end{figure}

\begin{figure}
\centering
\includegraphics[width=\columnwidth]{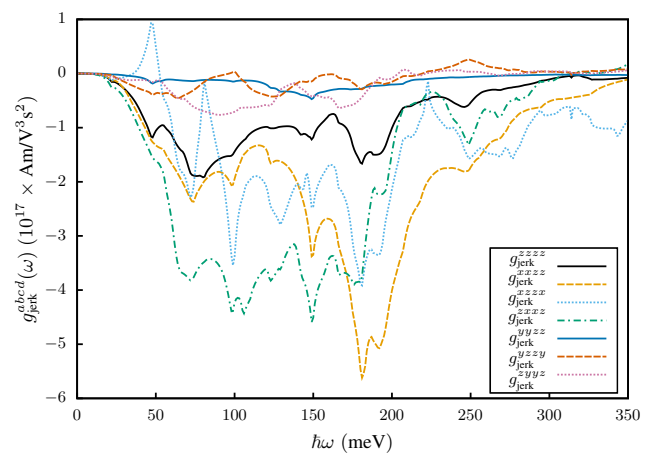}
\caption{Spectra of the remaining nonvanishing components of the jerk-current tensor that were not included in Fig. \ref{fig:jerk_spr}.}
\label{fig:jerk_other}
\end{figure}

In this appendix we provide further information regarding the agreement between the \textit{ab-initio} and the Wannierized bands and show further nonzero components of the jerk current tensor in Eq. \eqref{eq:jerk_tensor}. The quality of the wannierization procedure described in Sec. \ref{sec:methods} can be verified in Fig. \ref{fig:scf_vs_wann}, where we plot the \textit{ab-initio} bands obtained using the QUANTUM ESPRESSO package together with the bands obtained using the WANNIER90 package. In Fig. \ref{fig:jerk_other} we show the rest of the nonzero jerk current tensor components as a supplement to Fig. \ref{fig:jerk_spr}. Inspection of the figure reveals  that all components are smaller in magnitude than the dominant $g_{\text{jerk}}^{xxxx}$ component, and that 
the $g_{\text{jerk}}^{xxzz}$, $g_{\text{jerk}}^{zxxz}$ and $g_{\text{jerk}}^{xzzx}$ components are larger 
than the ``main axis" $g_{\text{jerk}}^{zzzz}$ component.

\bibliographystyle{apsrev4-2}
\bibliography{bibdata}

\end{document}